\documentclass[prb,amsmath,twocolumn,superscriptaddress,showpacs]{revtex4}

\newcommand{\dvec}[1]{\ensuremath{\mathbf{#1}}}
\newcommand{\p}{\dvec{p}}
\newcommand{\q}{\dvec{q}}
\renewcommand{\r}{\dvec{r}}

\begin{document}
\title{Electronic compressibility of graphene: The case of vanishing
electron correlations and the role of chirality}
\author{D. S. L. Abergel} 
\affiliation{Department of Physics and Astronomy, University of
Manitoba, Winnipeg, Canada R3T 2N2}
\author{Pekka Pietil\"ainen}
\affiliation{Department of Physical Sciences/Theoretical Physics,
University of Oulu, Oulu FIN-90014, Finland}
\author{Tapash Chakraborty}
\email{tapash@physics.umanitoba.ca}
\affiliation{Department of Physics and Astronomy, University of
Manitoba, Winnipeg, Canada R3T 2N2}
\begin{abstract}
A recent surprising finding that electronic compressibility measured
experimentally in monolayer graphene can be described solely in terms of
the kinetic energy [J. Martin, et al., Nat. Phys. \textbf{4}, 144
(2008)] is explained theoretically as a direct consequence of the linear
energy dispersion and the chirality of massless Dirac electrons. For
bilayer graphene we show that contributions to the compressibility from
the electron correlations are restored.  We attribute the difference to
the respective momentum dependence of the low-energy band structures of
the two materials.
\end{abstract}
\maketitle

In an interacting electron system of uniform density, the (inverse)
electronic compressibility $\kappa^{-1} \propto \partial\mu/\partial n$
(where $\mu$ is the chemical potential and $n$ is the electron density)
is a fundamental physical quantity that is intimately related to the
strength of interelectron interactions \cite{pines,mahan}. First
measured in 1992, the compressibility of a two-dimensional electron gas
\cite{jim} provides valuable information about the nature of the
interacting ground state, particularly in the strong-coupling regime
where (in addition to the exchange energy) the Coulomb interaction
energy is also known to play a dominant role. In this context, a recent
report on the measurement of electronic compressibility in monolayer
graphene \cite{monolayer} revealed behavior which was totally unexpected
\cite{martin}.  In this work, scanning single-electron transistor
microscopy was used to measure the change of local electrostatic
potential (and thereby change in local chemical potential) of a graphene
sample when the carrier density was modulated \cite{martin}. The
observed results for the local inverse compressibility were found to be
quantitatively described by the kinetic energy alone and the authors
speculated that the exchange and correlation energy contributions to the
compressibility either each other cancel out or are negligibly small.
This interesting puzzle has remained unsolved because the approximate
theoretical schemes adopted by various authors to investigate electron
correlations in graphene \cite{polini,RG} do not find any such
cancellations. Similarly, the recently reported Hartree-Fock studies of
compressibility \cite{HF} in monolayer and bilayer graphene do not
consider electron correlations at all and are therefore not in a
position to address this important issue. 

In this paper, we investigate the role of electron correlations in
monolayer and bilayer graphene. We show how in monolayer graphene, two
fundamental properties of the system, viz., the linear energy
dispersion and chirality conspire to allow the exchange and correlation
contributions to vanish, just as was observed in the experiment
\cite{martin}.  In bilayer graphene on the other hand, where
the low-energy quasiparticles are massive chiral fermions
\cite{mccann,bilayer}, the parabolic dispersion does not allow this
vanishing of the two energies, and the kinetic energy retains a
dependence on the electron correlation function which manifests in the
electron compressibility.

The low-energy charge carriers in monolayer graphene behave as massless
Dirac fermions described by the single-particle Hamiltonian $\mathcal
T^{\mathrm{m}}_1 \propto \boldsymbol{\sigma}\cdot \hat{\p}_1$, which
is linear in momentum $\p$ and where the subscript refers to the
coordinate label of the electron on which it acts.  The eigenstates of
the Hamiltonian are uniquely labeled by quantum numbers representing the
wave vector $\q=\p/\hbar$, the band (conduction/valence) $b$, valley
(pseudospin) $\xi$, and the $z$-component of the real electron
spin $\sigma$. The wave functions are of the form, $\psi(\r) = 
e^{i\q \cdot \r} \chi,$ where $\chi$ is an eight component spinor
\cite{ando}. A full analytical study of the many-electron system in
graphene is clearly an impossible task.  However, most of the clues to
the puzzle involving the measured compressibility described above can
be found at the level of two electrons, which is amenable to a fully
analytic solution. We therefore start with a two-electron system where
the electrons occupy the states $\alpha$ and $\beta$ corresponding to
the full sets of quantum numbers $(\q_{\alpha,\beta}, b_{\alpha,\beta},
\xi_{\alpha,\beta}, \sigma_{\alpha,\beta})$.  We denote by $\varphi$ the
antisymmetric noninteracting two-electron wave function
\begin{equation*}
	\varphi(\r_1,\r_2) = \frac{1}{\sqrt2}
	\left[ \psi^{}_\alpha(\r_1) \psi^{}_\beta(\r_2) 
	- \psi^{}_\beta(\r_1) \psi_\alpha(\r_2) \right].
\end{equation*}
The correlations due to the mutual Coulomb interaction are introduced by
multiplying the free-particle wave function by a generic correlation
factor $F$ as 
\begin{equation*}
	\Psi = F(\r_1,\r_2) \varphi(\r_1,\r_2).
\end{equation*} 
At this stage, a precise definition of $F$ is not necessary. The only 
requirements are that it should be a real function, and to preserve 
the antisymmtery of the correlated wave function $\Psi$ it is assumed 
to be symmetric with respect to exchange of the particle
coordinates, \textit{i.e.}, $ F(\r_1,\r_2) = F(\r_2,\r_1)$.

\begin{widetext}
In order to evaluate the two-particle energy we have to normalize the 
wave function $\Psi$. A straightforward calculation yields
\begin{equation*} 
	||\Psi||^2 = \langle\Psi|\Psi\rangle = \int d\r_1 \, d\r_2 \, F(\r_1,\r_2)^2 
	\left\{ 1 - \tfrac{1}{2} \delta_{\xi_\alpha \xi_\beta} 
	\left[1 + b^{}_\alpha b^{}_\beta \cos(\theta^{}_\alpha-\theta^{}_\beta)\right]
	\cos Q \right\},
\end{equation*} 
where $Q=(\q^{}_\beta - \q^{}_\alpha) \cdot (\r_1 - \r_2)$, and 
$\theta_{\alpha,\beta}$ are the polar angles of the momenta $\q_{\alpha,\beta}$. 
Evaluation of the expectation value of the kinetic part of the
Hamiltonian
$T^\mathrm{m} = \langle\Psi | \mathcal{T}^\mathrm{m}_1 
+ \mathcal{T}^\mathrm{m}_2 |\Psi\rangle$ leads to the expression 
\begin{align*} 
	T^\mathrm{m} &= \tfrac{1}{2} \hbar v^{}_F 
		\int\, d\r_1 \, d\r_2 \, 
		F \bigg\{ \tfrac{1}{4} i \delta_{\sigma_\alpha\sigma_\beta}
		\delta_{\xi_\alpha\xi_\beta}
		\Big[ e^{iQ} Z_{\alpha\beta}^1 
		\left(1 + b_\alpha b_\beta e^{-i(\theta_\beta - \theta_\alpha)} \right) 
		+ e^{iQ} Z_{\beta\alpha}^2 \left(1 
		+ b_\alpha b^{}_\beta e^{i(\theta^{}_\beta-\theta_\alpha)} \right) \\ 
	& \qquad\qquad \qquad\qquad \qquad\qquad + e^{-iQ} Z_{\beta\alpha}^1 
		\left( 1+ b_\alpha b^{}_\beta e^{i(\theta^{}_\beta-\theta_\alpha)} \right) 
		+ e^{-iQ} Z_{\alpha\beta}^2 
		\left(1 + b_\alpha b_\beta e^{-i(\theta_\beta - \theta_\alpha)} \right)
		\Big] \\ 
	&\qquad\qquad  - {i b^{}_\beta} \bigg( \cos\theta^{}_\beta
		\frac{\partial F}{\partial x_1} + \cos\theta^{}_\beta
		\frac{\partial F}{\partial x_2} + \sin\theta^{}_\beta
		\frac{\partial F}{\partial y_1} + \sin\theta^{}_\beta
		\frac{\partial F}{\partial y_2} + 2i q^{}_\beta F \bigg)\\
	&\qquad\qquad - {i b_\alpha} \bigg( \cos\theta_\alpha
		\frac{\partial F}{\partial x_1} + \cos\theta_\alpha
		\frac{\partial F}{\partial x_2} + \sin\theta_\alpha
		\frac{\partial F}{\partial y_1} + \sin\theta_\alpha
		\frac{\partial F}{\partial y_2} + 2iq_\alpha F \bigg)
		\bigg\},
\end{align*} 
where $F$ stands for $F({\bf r}_1,{\bf r}_2)$, $v_F$ is the Fermi
velocity, and $Z_{\alpha\beta}^j$ is shorthand for 
\begin{equation*}
	Z_{\alpha\beta}^{j} = 
	\left( b^{}_\beta e^{i\theta^{}_\beta} 
		+ b_\alpha e^{-i\theta_\alpha} \right)\frac{\partial
		F}{\partial x_j} 
	- i\left( b^{}_\beta e^{i\theta^{}_\beta} 
		- b_\alpha e^{-i\theta_\alpha} \right) \frac{\partial
		F}{\partial y_j} 
	+ i q^{}_\beta F\left(b^{}_\beta 
		+ b_\alpha e^{i\left(\theta^{}_\beta-\theta_\alpha\right)}\right).
\end{equation*}
\end{widetext}
Due to the linearity of $\mathcal{T}^\mathrm{m}$ in the momentum
operators, only the first-order derivatives appear in the integrand. 
Terms in $T^\mathrm{m}$ of the form 
\begin{equation*}
	\int d\r_j \, F \frac{\partial F}{\partial x_j} =\frac{1}{2} \int d\r_j
	\frac{\partial}{\partial x_j}F^2, 
\end{equation*}
clearly vanish due to the antisymmetry of the integrand. Most of the 
terms left after the volume integration cancel each other as a consequence 
of the spinor structure of the single-particle wave functions, which is
a direct manifestation of the chirality of the electrons. The only
surviving terms sum to \begin{equation*}
	T^\mathrm{m} = \hbar v^{}_F 
	(b_\alpha q_\alpha + b_\beta q_\beta) ||\Psi||^2, 
\end{equation*}
that is, the kinetic energy 
expectation value $\langle\mathcal{T}^\mathrm{m}\rangle$ is simply the sum 
of the single free-particle kinetic energies, 
\begin{equation*}
	\langle{\mathcal T^\mathrm{m}}\rangle 
	= \frac{T^\mathrm{m}}{||\Psi||^2} 
	= \hbar v_F(b_\alpha q_\alpha+b^{}_\beta q^{}_\beta) 
	= T_0^{\mathrm{m}}
	=\frac{\langle\varphi|{\mathcal T^\mathrm{m}}|\varphi\rangle}{||\varphi||^2} 
\end{equation*}
and does not depend on the correlation function $F$ at all. We expect 
similar cancellations for higher electron numbers, although analytical 
expressions become intractable already at the level of three electrons.

Complete cancellation of correlation contributions to the kinetic 
energy (never observed in conventional electron systems) creates 
an unusual situation as we shall now describe. In the thermodynamic limit, 
the potential energy (per particle) $\mathcal{V}$ is usually
expressed in the form 
\begin{equation*}
	\langle{\mathcal V}\rangle = n\int d\r \, [g(r)-1] V_{\rm Coul}(r),
\end{equation*}
where $n$ is the single particle number density, $V_{\rm Coul}$ is the 
Coulomb potential and $g(r)$ is the pair-correlation function which, for 
$\r = \r_1 - \r_2$, is given by
\begin{multline*} 
	g(|\r|) = \frac{N(N-1)}{||\Psi||^2n^2} \times \\
	\times\int d\r_3 \ldots d\r_N \, |\Psi(\r_1,\r_2,\r_3,\ldots,\r_N)|^2, 
\end{multline*}
where $N$ is the total number of electrons. The energy functional (per
particle) ${\mathcal E}^\mathrm{m}$ is now 
\begin{equation*}
	{\mathcal E}^\mathrm{m} 
	= t_0^\mathrm{m} + n \int d\r\, [g(r)-1] V_{\mathrm{Coul}}(r), 
\end{equation*}
where $t_0^\mathrm{m} = T_0^\mathrm{m}/N$ is the kinetic energy per
particle. Its variation with respect to $g(r)$, an essential step in
determining the optimal $g(r)$, 
would yield an unusual Euler-Lagrange equation, $V_{\mathrm{Coul}}(r)=0,$ 
which is clearly not the case in graphene \cite{note}. To resolve this 
dilemma we note that the energy functional ${\mathcal E}^\mathrm{m}$ 
is actually not bounded below: we can choose correlations such that the
potential energy takes arbitrarily large negative values. This implies
that to determine the optimal $g(r)$ the energy functional derived above
is not sufficient and additional physical constraints, for example, that
$g(r)$ should correspond to the correct number of states in each band,
would be necessary. Clearly, determination of the optimal
pair-correlation functions for massless Dirac fermions in graphene is a
nontrivial problem \cite{chandre}.  However, we believe that the
expression for the functional ${\mathcal E}^{\mathrm{m}}$ is of the
correct form; i.e., once the correct pair-distribution function
$g(r)$ is found, one could evaluate the correct energy from the above
form of the energy functional.

Let us now turn our attention to the compressibility as defined in the
introduction.  To that end we first evaluate the variation
$\delta_n{\mathcal E}^\mathrm{m}$ of $\mathcal E^\mathrm{m}$ 
with respect to $n$: 
\begin{multline*}
	\delta_n{\mathcal E}^\mathrm{m} 
	= \frac{\partial t_0}{\partial n}\delta
		n + \delta n \int d\r \, [g(r)-1] V_{\mathrm{Coul}}(r) \\  
	+ n\int d\r \, V_{\mathrm{Coul}} \frac{\delta g(r)}{\delta n}\delta n.  
\end{multline*} 
From this we can read the derivative as 
\begin{equation*}
	\frac{\partial \mathcal{E}^\mathrm{m}}{\partial n}
	= \frac{\partial t_0}{\partial n} + \int d\r \, \left[ 
                g(r) - 1 \right] V_{\mathrm{Coul}} 
	+ n \int d\r \, V_{\mathrm{Coul}} \frac{\delta g(r)}{\delta n}. 
\end{equation*}
The compressibility will then be proportional to 
\begin{multline*}
	\frac{\partial^2{\mathcal E}^\mathrm{m}}{\partial n^2} = \frac{\partial^2 t_0}{\partial n^2}  
		+ 2\int d\r \, V_{\mathrm{Coul}}\frac{\delta g(r)}{\delta n} \\
	+ n \int d\r \, V_{\mathrm{Coul}} \frac{\delta(\delta g(r)/\delta n)}{\delta n^2}.  
\end{multline*} 
We therefore need to make an assumption or estimation of the functional
derivative $\delta g(r) / \delta n$. The pair-correlation function is a
many-body quantity, so its exact evaluation is impossible. Also,
it is not possible to calculate functional derivatives numerically, and
any analytical approximation will necessarily obscure the true
functional dependence that we require.
In conventional two-dimensional electron systems, $g(r)$ varies only
slightly as a function of density except at very low densities where it
starts to develop a prominent peak as a precursor to Wigner
crystallization \cite{ceperley}.  
Therefore, we appeal to previous work which shows that there are no
phase transitions (such as Wigner crystallization) as the density of the
graphene system is varied \cite{wigner} and so we expect that the
functional variation of $g(r)$ with the density will be negligable in
this system.
Alternatively, we could consider a slightly less stringent condition
$\int d\r \, V_{\mathrm{Coul}} \frac{\delta g(r)}{\delta n} = 0$, which
implies that the interaction energy depends linearly on the density of
Dirac electrons. 
Under either of these assumptions, the compressibility is described
entirely by the kinetic energy
\begin{equation*}
\frac{\partial^2{\mathcal E}^\mathrm{m}}{\partial n^2}=\frac{\partial^2 t_0}{\partial n^2}, 
\end{equation*}
in accordance with the experimental observation \cite{martin}. In
arriving at this striking result, there are two basic properties of
monolayer graphene that play crucial roles: the linear energy dispersion
and chirality of massless Dirac electrons. 

This immediately invites the question: What happens in bilayer graphene
\cite{mccann,bilayer}, where the low-energy charge carriers behave as 
massive chiral fermions and as such the Hamiltonian is quadratic in 
momentum operators near the charge neutrality point. 
The single-particle Hamiltonian is $\mathcal{T}^\mathrm{b}_1 =
-\tfrac1{2m^*} (\boldsymbol{\sigma}\cdot\hat{\p}_1) \sigma_x
(\boldsymbol{\sigma}\cdot\hat{\p}_1)$ where $m^*$ is the effective
electron mass generated by the inter layer coupling, and has the
spectrum $\epsilon_\alpha = (\hbar q_\alpha)^2$ associated with it.
Evaluating the expectation value of an arbitrary two-particle wave
function as in the monolayer case, an intermediate expression for the
kinetic part of the two-particle energy is
\begin{widetext}
\begin{align*}
	T^\mathrm{b} &= \tfrac12(\epsilon_\alpha + \epsilon_\beta) 
	- \int d\r_1 d\r_2 \, \frac{\hbar^2F}{4m^*} e^{-iQ} 
		\Big\{iq_\beta \left( \left[ \cos\theta_\beta 
		+ \cos(2\theta_\alpha-\theta_\beta) 
                  \right] \frac{\partial F}{\partial x_1} 
		+ \left[ \sin\theta_\beta + \sin(2\theta_\alpha
                - \theta_\beta) \right] 
	          \frac{\partial F}{\partial y_1} \right) \\
	& \qquad\qquad \qquad\qquad\qquad\qquad\qquad 
	+iq_\alpha \left( \left[ 
                 \cos\theta_\alpha + \cos(\theta_\alpha-2\theta_\beta)
                 \right] \frac{\partial F}{\partial x_2}
		+ \left[ \sin\theta_\alpha - \sin(\theta_\alpha-2\theta_\beta) 
                  \right] \frac{\partial F}{\partial y_2} \right) \Big\} \\
	& \qquad - \int d\r_1 d\r_2 \, \frac{\hbar^2F}{4m^*} e^{iQ}
		\Big\{iq_\alpha \left( \left[ \cos\theta_\alpha 
		+ \cos(\theta_\alpha-2\theta_\beta)\right] 
                  \frac{\partial F}{\partial x_1}
		+ \left[ \sin\theta_\alpha - \sin(\theta_\alpha-2\theta_\beta) 
                  \right] \frac{\partial F}{\partial y_1} \right) \\
	& \qquad \qquad\qquad\qquad\qquad + iq_\beta \left( 
                  \left[\cos\theta_\beta + \cos(2\theta_\alpha-\theta_\beta)
                  \right] \frac{\partial F}{\partial x_2} 
		+ \left[ \sin\theta_\beta + \sin(2\theta_\alpha-\theta_\beta) 
                  \right] \frac{\partial F}{\partial y_2} \right)
		\Big\}
\end{align*}
\end{widetext}
where we have already excluded terms containing mixed second derivatives
of $F$ which are identically zero on integration, and those which
trivially sum to zero. The integrals of terms with single derivatives of
$F$ are finite, and the prefactors (coming from the pseudospinor part of
the products of wave functions in the expectation value) do not cancel
each other as they did in the monolayer case, but constructively sum to
a finite result.  This noncancellation is a feature of the sublattice
structure of the electronic wave function in bilayer graphene, resulting
from the quadratic nature of the low-energy dispersion relation. On
evaluation of the remaining integrals, and after some elementary
algebra, the energy functional is found to be 
\begin{multline*}
	\mathcal{E}^\mathrm{b} 
	= t_0^\mathrm{b} + \left\langle \mathcal{V} \right\rangle 
	+  \frac{\hbar^2}{8m^*}
	\frac{ \tilde{F}^2 }{||\Psi||^2} \cos(\theta_\alpha-\theta_\beta) \\
	\times \left[ (q_\alpha^2 + q_\beta^2) 
        \cos(\theta_\alpha-\theta_\beta) - 2 q_\alpha q_\beta \right]
\end{multline*}
where $\tilde{F}$ is the Fourier transform of the correlation function.
We can easily see that there is a nonzero contribution from the
electron correlations to the kinetic energy in this functional, and
therefore taking the derivatives with respect to $n$ yields a
compressibility which depends nontrivially on them. 
It is clear that this additional term will also be present in the many
body energy, as it's integral over momentum is manifestly finite.
We would also expect that for bilayer graphene where the excess
electron density is high enough that the Fermi energy is in the energy
range where the linearity of the spectrum is restored, that the effect
of the correlations in the energy functional will again be supressed. 
Quantitative computation of this term requires precise knowledge of the
radial dependence of $F$, and the relation between $F(r)$ and $g(r)$
\cite{stevens}. Both of these issues are beyond the scope of the present
paper, however, as indicated in our present results for bilayer
graphene, experimental observation of a shift in compressibility from
the pure kinetic energy contribution would provide a way to directly
determine the strength of electron correlations in that system.

To conclude, we have demonstrated that in monolayer graphene, the
electron correlations analytically vanish from the two particle kinetic
energy. This, and the negligability of $\delta g(r) / \delta n$ lead to the
absence of the electron correlation function in the compressibility, as
seen in recent experiments \cite{martin}. 
Conversely, the restoration of a quadratic band structure in bilayer
graphene means that the correlations are present in the kinetic energy
functional and compressibility in this case. 
Our work strongly suggests that quantitative agreement between the
single particle theory and the experimental results in monolayer
graphene has its origin in the fundamental properties (and in
particular, the linear band structure) of this system.
Also, the experimental system exhibits nonhomogeneity of the charge
distribution, so a full many-body calculation of the compressibility
would have to include this detail. However, the cancellation of the
correlation function from the monolayer two particle energy
functional is independent of the energy of the electrons, and will
therefore persist in the inhomogeneous system.

The work was supported by the Canada Research Chairs Program and the NSERC 
Discovery Grant.

\end{document}